# An Empirical Evaluation of AriDeM using Matrix Multiplication

**Patrick Mukala**

Department of Computer Science, Faculty of ICT, Tshwane University of Technology

**ABSTRACT**

For a long time, the Von Neumann has been a successful model of computation for sequential computing .Many models including the dataflow model have been unsuccessfully developed to emulate the same results in parallel computing. It is widely accepted that high performance computation is better-achieved using parallel architectures and is seen as the basis for future computational architectures with the ever-increasing need for high performance computation. We describe a new model of parallel computation known as the Arithmetic Deduction Model (AriDem) which has some similarities with the Von Neumann. A theoretical evaluation conducted on this model in comparison with the predominant von Neumann model indicated AriDeM to be more efficient in resources utilization. In this paper, we conduct an empirical evaluation of the model and the results reflect the output of the theoretical evaluation.

## 1. INTRODUCTION

The purpose of this paper is to evaluate AriDeM by establishing a systematic comparison with the von Neumann model. Making use of matrix multiplication as our case study, we developed code to simulate the behavior of the AriDeM architectural model and execute related programs to analyze the von Neumann model. An interpretation of these results for both models follows.

One aspect of this evaluation is to verify that it is indeed possible to express and evaluate programs using AriDeM or the element model. In order to perform this empirical evaluation, the matrix multiplication case study is used as a benchmark to provide some indications in terms of the performance. An analysis is made as to how this element model compares to a von Neumann model using matrix multiplication.

## 2. THE ARITHMETIC DEDUCTION MODEL (AriDeM)

AriDeM stands for Arithmetic Deduction Model. It is an architectural model based on natural rules of computation. Despite the predicted better performance compared to other approaches, AriDeM is similar to the von Neumann model in many ways. However, instead of processing instructions it processes elements. More on this can be found in [3]. For this reason, in this paper we refer to AriDeM as the element model and the von Neumann model as the instruction model.

### 2.1. Main Concepts of AriDeM

At the heart of the model is an element. An element is made up of three parts including an identifier, indices and a value as shown in Table 1. For example, *length = 5cm* is an element.

Table 1: Element description

| Field | Description | Comment |
|---|---|---|
| identifier | binary number | used to identify relations |
| indices | list of binary numbers | ensures unique meaning |
| value | one or two binary numbers | can be a single or tuple value |

Since an element refers to relations (also referred to as relationships), the identifier points to a list of relations as described in table 2.

Table 2: Description of relations

| Field | Description | Comment |
|---|---|---|
| Identifier | Binary number | Used for the identifier of the new element |
| Operation | Binary number | Specifies the operation required for the new element |
| parameters | binary numbers | used by operation |

For example, *square_area = length²* is a relation. An expression of how these two concepts relate is shown in figure 1. AriDeM expresses through Figure 1 that mappings can describe relations between sets in the problem domain and given an element in the domain of one of these relations enables a deduction to be made about an element in the range. For example given the relation

$$square\_area = length^2$$

and the element

$$length = 5cm$$

one can deduce based on the given relation,

$$square\_area = 25cm^2$$

This means that when an element like *length = 5cm* comes into existence, it is processed by applying it to relations such as *square_area = length²*. The element together with the relation is then used to deduce another element and in this case, *square_area =25cm²*. Once the element has been processed, it is discarded and the newly created element can be applied to other relations that possibly create new elements.

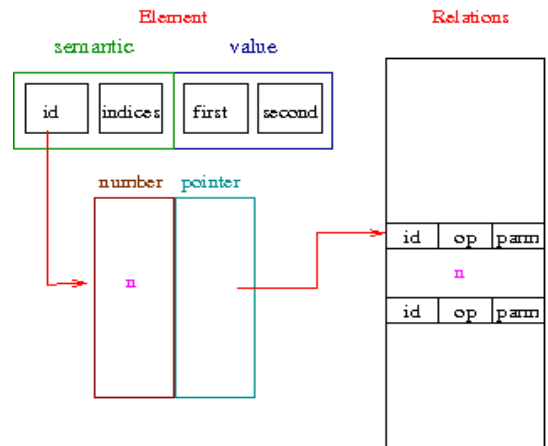

Figure 1: Interplay between element and relations

## 2.2. Basic principles

The elements consist of information that uniquely conveys the meaning of the data as well as a value. The information is used to determine how the element is processed. Processing an element results in the creation of zero or more elements. The program execution completes when there are no more elements to process. The execution cycle for both the instruction and element models is as follows:

Table 3: Instruction vs. element model execution cycle

| Instruction Model | Element Model |
|---|---|
| Get instruction | Pop Element |
| Get Data | Get relation |
| Perform op | Perform op |
| Store result | Push result |

To illustrate the mechanism we start with a simplified version of the model. The relationships are expressed as a definition of an element e.g. $a = -b$. This can also be written as $b \rightarrow a$, which is an alternative AriDeM notation for a relation. The elements are expressed as a value associated with an identifier e.g. *b = 5*. In this example the element *b = 5* starts on the stack of unprocessed elements.

The first step of the cycle is to pop *b = 5*. The second step is to use the identifier of the element to get the relation b → a. The last step is to compute −b or −5 to create the element *a = −5* and store it on queue of unprocessed elements.

## 3. EXPERIMENTATION

To determine the behavior of the element model, this experimentation looks at how effectively the processing load can be distributed across the different processors in the architecture as well as the duration of the execution. To achieve this, the experiment explores the configuration for the simulation of this model in C using openmp. Although the simulation does not provide a complete accurate performance of the model, the purpose is to get an approximate indication of the features and behavior of the model.

The results of the programs for both the von Neumann and AriDeM models give an indication of how these two compare and validate the observations made during theoretical evaluation. For the AriDeM model, these results help detect any weaknesses in the model, set a platform for fine tuning some of the design concepts of the model, and more importantly determine how effectively the distribution of workload occurs as well as obtaining a clear indication of the performance of the model in terms of elements created and messages passing. This means the analysis involves looking at the number of messages sent for a matrix multiplication of size n, the number of instructions (elements) to process a message by all involved processors. This chapter also looks at an analysis of how the problem scales with increasing size of n.

To achieve all of this, during this empirical study just like in the previous chapter, the two models are evaluated on two levels: number of processors and size of matrix. The codes used to perform this evaluation for each respective model are run on a smaller number of processors (2) and gradually increases the number of processors as well as the size of matrix in order depict the scalability in both models. A conceptual view of the operation behind the entire execution process for the element model can be seen in figure 5. The C program code we developed was executed in order to verify the results.

## 4. EVALUATION AND RESULTS

In order to get the results needed for this evaluation, we considered four different sizes of matrix executed on different number of processors. The sizes of matrix considered were a 40 by 40 matrix (40 x 40), a 60 by 60 matrix (60 x 60), an 80 by 80 matrix (80 x 80) as well as a 100 by 100 matrix (100 x 100).

Concerning the number of processors used for performance testing, we considered increasing the number of processors from 2 to 16. In practical, we executed the programs on 2 processors, 4 processors, 8 processors and 16 processors. These processors are randomly picked from the SUN cluster. We requested 20 processors on the SUN cluster from the Centre for High Performance Computing (CHPC) as computational resources to avail for our experimentation and testing.

Another important point to note at this stage is that the results reflected in this experimentation were not obtained during the execution. In order to get more precise results, the execution time was recorded after running the scripts at least 10 times. This means that we had to submit the same script (as shown in figure 9) 10 times and return the average time when applicable to ensure that we have more accurate values.

### 4.1 Description of Codes

#### 4.1.1 Instruction model (von Neumann)

MPI was used to write a parallel program using C. The fundamental underling explanation of the code used to test the instruction model is that in the course of executing this code, data in matrix A is copied to every slave (processor) and data in matrix B is divided into blocks and distributed among the different processors or slaves. During execution, the data is distributed across the slave-processors in order to perform the actual matrix multiplication and each slave sends the corresponding results to the master.

First, the program contains code to generate random numbers to be used for computation. Two *for loops* are used to ensure that the matrices are evenly populated with random numbers. This is performed by the master-processor. The master also ensures the coordination and distribution of instructions load to the remaining slaves (processors). Each slave performs its instructions and sends the results back to the master for coordination.

Each slave receives instructions pertaining to its tasks. This means, the slave receives instructions about what to do and how to do it. The sequence in which instructions are executed is important in order to carry out the computational task successfully. If an instruction is not properly receives, it needs to be resent. This logically causes a delay in the execution of the program because results produced by one or more slaves may depend on the results generated by other slaves.

The code also expresses the communication between the Master and the different Slaves. Slaves request instructions from the Master and the later sends requested information based on the sequence of data execution and available results.

### 4.1.2 Element Model

Lee and his co-workers emphasize that research in modern computer architecture heavily relies on simulation techniques [1]. This is because conducting a thorough evaluation of a new architectural model is a complex and time consuming process. Hence, simulation is very important and indispensable in computer architecture research.

As further noted in [1], architectural simulators are widely used in computing to help identify performance bottlenecks, analyze interactions among different hardware and software components and lastly help evaluate the impact of the projected architecture on system performance. Thus, we used the software simulated architecture for AriDeM to perform matrix multiplication.

This is critical to grasp at this stage as supported by Wunderlich and colleagues [2]. According to their research findings, 1 ms of execution in real time can correspond to days of simulation time [1,2] .Since we are using the simulator for AriDeM, the reflected results are much higher than the results for a real time execution. However, these results are helpful as they give us an indication on the performance of the architecture in terms of scalability.

Approximately nine files contain the code needed to make the simulator fully operational. In the AriDeM code, the master ensures the coordination and distribution of elements to available slaves throughout the entire computation process.

Firstly, random numbers are generated in order to ensure we have data to feed to the processors for computation. The master-processor first reads in through all elements and distributes them to slave-processors by looping through all of them until none of the slaves is idle. The master checks for available processors and ensures that elements are assigned and distributed to these slaves accordingly. The slaves on the other hand, read available mappings and send requests to the master. Looping through elements, these slaves process the received elements and continually send request for available partial or full elements in order to perform the computation. This makes each slave active (not idle) until there are no more elements to be processed.

Other files include codes used to cater for mappings, hashing, printing as well as semantic processing. Each element has a semantic attached to it, and based on its unique reference number the semantic will help determine how the element will be used by the receiving slave-processor.

In order to generate results, we needed to write different scripts to be submitted to the SUN cluster. A typical script would contain relevant information such as number of processors required for a particular execution, time...as shown in Figure 9 below.

**Script used to submit job on cluster for execution**

```
#!/bin/sh
#MSUB -l nodes=10:ppn=2
###MSUB -l partition=ALL
#MSUB -l walltime=00:00:04
#MSUB -m be
###MSUB -V
#MSUB-o /export/home/pmukala/scratch/Final3/dix/dix2.out
#MSUB-e /export/home/pmukala/scratch/Final3/dix/dix2.err
#MSUB -d /export/home/pmukala/scratch/Final3/dix
#MSUB -mb
#MSUB -M patrick.mukala@gmail.com
/opt/gridware/sun-hpc-ct-8.2-Linux-gnu/bin/mpirun -np 2 dix
```

Figure 2: Sample of scripts used to submit the codes on SUN cluster

In the script in Figure 2, the second line *MSUB –l* just specifies the requested number of nodes as well as the corresponding number of processors per node. *MSUB –m* simply specifies the estimated maximum time that the program will take to execute. The lines from *MSUB –o to MSUB –d* help specify the output files to be generated after the program is executed. And the last line */opt/gridware/sun-hpc-ct-8.2-Linux-gnu/bin/mpirun -np 2 dix* specifies the command to use in order to execute the program. In this line, *mpirun* simply allows the program to run following the inherent options *–np* depicts the number of processors which is 2 and the last word *dix* indicates the executable file to be used.

In the next section, we look at the actual empirical data for both models. This information is structured according to how the time varies with the increasing number of processors, the execution time in light with the increasing size of matrices and the number of instructions executed versus the number of processed elements for each size of the matrices.

### 4.2 Time vs Number of Processors

It is important to note that the point here is not to directly determine whether AriDeM will execute the same task in less time than von Neumann will. We explained previously that the results in terms of execution time generated with the AriDeM simulator are higher than the possible results that can be generated in real time execution. Hence, the time metric used here will help us determine the scalability of both models. Note that the time is expressed in milliseconds.

Table 1
Von Neumann (Indication of time (ms) per matrix per processors)

|         | 2 p    | 4 p   | 8 p    | 16 p   |
|---------|--------|-------|--------|--------|
| 40 x 40 | 4.72   | 2.56  | 3.409  | 3.2113 |
| 60 x 60 | 10.058 | 6.466 | 10.8121| 7.774  |
| 80 x 80 | 20.794 | 9.016 | 14.717 | 16.559 |
| 100x100 | 73.357 | 15.62 | 21.03  | 24.272 |

Table 2
AriDeM (Indication of time (ms) per matrix per processors)

|          | 2 p      | 4 p      | 8 p      | 16 p     |
|----------|----------|----------|----------|----------|
| 40 x 40  | 88.733   | 70.086   | 324.738  | 25.423   |
| 60 x 60  | 826.578  | 315.925  | 607.044  | 292.929  |
| 80 x 80  | 1280.438 | 379.082  | 209.524  | 255.179  |
| 100 x100 | 5515.629 | 3591.148 | 1461.293 | 1095.009 |

**Observation 1:**

These two tables give a general view of the performance of the two models based on these two parameters: number of processors and size of matrices. Note that both models are able to scale as the number of processors increases. However, as shown in figure 3 and figure 4, AriDeM shows indicate of a better performance as it scales a little bit more effectively than von Neumann does. Especially with bigger sizes of matrices, for example with the 100 by 100 matrix, AriDeM scales perfectly with the time decreasing as the number of processors increases. Unlike AriDeM, von Neumann shows a bit of discrepancy with the largest number of processors taking more time than smaller number of processors as shown in Figure 3 and 4 below.

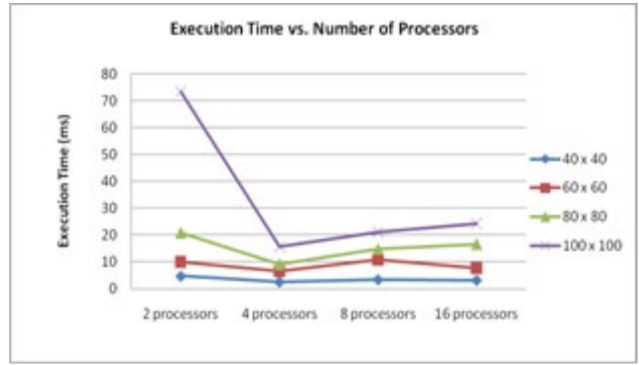

Figure 3: Instruction model. Graph indicating time (ms) per matrix size and number of processors

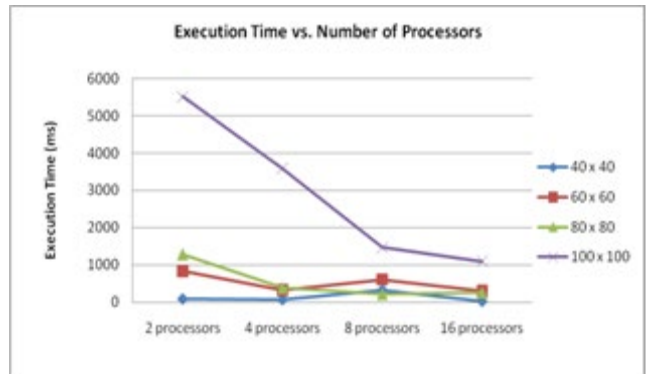

Figure 4: Element model. Graph indicating time (ms) per matrix size and number of processors

**Observation 2**:

These two graphs show an indication of how scalable these two models are. Let's note that for both models, it is clear that more time is needed to execute bigger size of matrices and that logically, less time will be needed. However, this is not entirely the case especially for the instruction model. Looking at the graph in figure 10, we can see some discrepancies in terms of execution time and increasing size of matrix. We notice that as the size of matrix increases, the model does not strongly express that it is scalable. On the contrary, when the matrices 80 by 80 and 100 by 100 even 60 by 60 are run on 8 and 16 processors, we notice that the time actually goes up with 16 processors instead of decreasing. The same slightly happens for AriDeM but only with the 60 by 60 matrix. Otherwise, the performance of AriDeM indicates that it slightly more scalable than von Neumann in this case.

Under normal circumstances, both models are expected to be entirely scalable. Since our testing platform is a cluster, there are some factors that influence these results. Due to the configuration of the cluster containing interconnected nodes, three factors have likely affected the final results. These factors include computation, idle and communication time.

The computation time is the only time that we have tried to measure for each model and each case. This represents the time taken in executing the instructions in order to perform the matrix multiplication. This implies that we look at how long it took the architecture to actually fetch data in each of the related matrices, perform the matrix multiplication and store the solution in the resulting matrix. Having this information for both models would help determine their performance bottlenecks. However the results we got were also affected by idle time.

The SUN cluster used for testing has a von Neumann architecture. Hence, all these factors affecting the final results for both the instruction and element models are to some extent influenced by the the underlying architecture of the system. The idle time is hence caused by the inability of the instruction model to effectively distribute data or send instructions to processors.

Unlike the von Neumann model, AriDeM caters for this by enforcing data parallelism.
In the element model, the master loops through all slaves during processing and sends elements to slaves as soon as they are free and this eliminates any idle time. Idle time causes delays in processing time, if one processor has to wait for results from other processors to execute its instructions, this will definitely impact on the overall performance and execution time. AriDeM on the other hand enforces independency among processors during processing. Each processor processes its elements and sends requests once all the elements are processed and the execution process stops when there is no element on the queue. And this is when all the processors become idle.

Communication time is also another factor that has a major impact on the generated output. Because the processors are dependent on one another, the communication interval, as explained in our literature survey, causes the delay in processing time. If processors have to exchange data and communicate frequently, then this increases the inherent computation time and has an impact on the architecture performance.

Other factors might as well affect the performance of the architecture. One is the costs related to unaccounted –for overhead. Issues such as load imbalances may cause this overhead. Meaning that there are computation imbalances among processors where some processors execute more instructions than others do. Another issue is replicated computation, this is due to events such as ignoring to parallelize portion of the programs. One last issue causing overhead is competition for bandwidth among processors. Each processor wants to use as high bandwidth as possible and this causes increased communication costs.

Another underlying factor is network traffic since the cluster is a network of nodes. The processors used for testing are located on different nodes, this sometimes causes delays because of memory limitations, and network traffic.These factors make it difficult and complex to obtain scalable computation on a cluster. Next, we consider the changes in time as we increase the size of the matrices

### 4.3 Time vs increasing size of matrix

Once again, the data in tables 3 and 4 as well as the graphs in figures 4 and 5 help us to have a view of how scalable the two models are. As we increase the size of matrices, we get an indication of how do these models scale with the changing number of processors as shown in figures 4 and 5 for the instruction and element models respectively. Note that the time is expressed in milliseconds.

Table 3
Von Neumann (Indication of time (ms) per processor per matrix)

|  | 40 x 40 | 60 x 60 | 80 x80 | 100 x100 |
| --- | --- | --- | --- | --- |
| 2 processors | 4.72 | 10.058 | 20.794 | 73.357 |
| 4 processors | 2.56 | 6.466 | 9.016 | 15.62 |
| 8 processors | 3.409 | 10.8121 | 14.717 | 21.03 |
| 16 processors | 3.2113 | 7.774 | 16.559 | 24.272 |

Table 4
AriDeM (Indication of time (ms) per processor per matrix)

|  | 40 x 40 | 60 x 60 | 80 x 80 | 100 x100 |
|---|---|---|---|---|
| 2 processors | 88.733 | 826.578 | 1280.438 | 5515.629 |
| 4 processors | 70.086 | 315.925 | 379.082 | 3591.148 |
| 8 processors | 324.738 | 607.044 | 209.524 | 1461.293 |
| 16processors | 25.423 | 292.929 | 255.179 | 1095.009 |

**Observation 3:**

These two tables contain almost the same information as in tables 1 and 2. The only difference here is that we changed columns just to draw graphs that will help us get a different view of the scalability of the two models. The graphs in figures 4 and 5 indicate that as the size of matrices increases AriDeM becomes more scalable. Unlike the instruction model, the element model is able to perform better with bigger sizes as we increase the size of matrices despite all the factors affecting performance we discussed earlier.

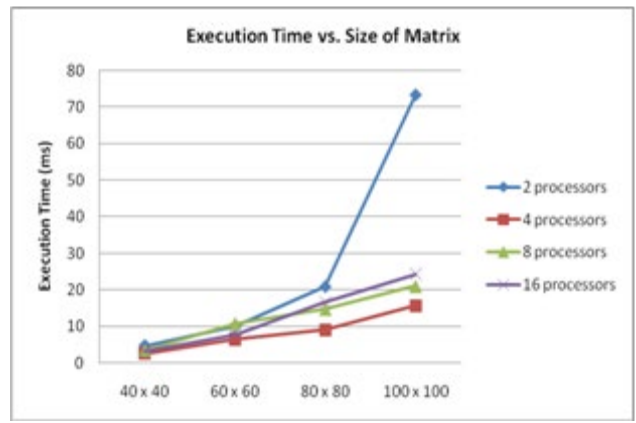

Figure 4: Von Neumann (Graph indicating time (ms) per processors per matrix)

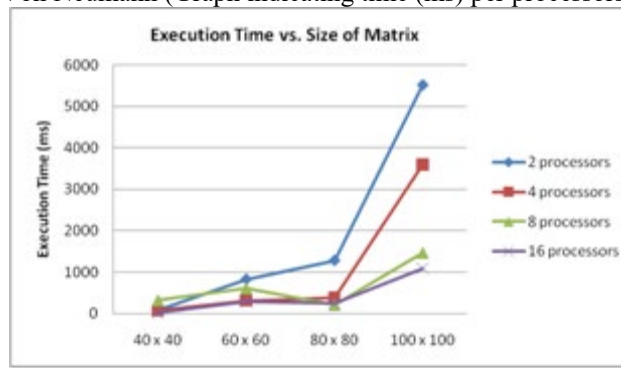

Figure 5: AriDeM (Graph indicating time (ms) per processors per matrix)

**Observation 4:**

We can once again have a definite picture of how scalable the models are by looking at graphs in figures 4 and 5 for both the instruction and element models respectively. As the graphs indicate, it is clear that more time is needed to execute bigger size of matrices and that logically less time will be needed when the number of processors increases. The instruction model has problems showing this correctly due to previously discussed factors affecting its performance. The element model however manages to show some signs of better scalability than the instruction model especially with bigger sizes of matrices.

The time metric helped determine how scalable both models are and what the related performance bottlenecks could be identified. The intention of our evaluation was to also see if AriDeM could show clear an indication of a parallel model of computation. This has been determined by its scalability. The fact that AriDeM can even be compared to von Neumann and produced almost the same results and even better shows that more attention needs to be directed to the model.

Next we look at the number of instructions executed during processing by the instruction model as compared to the number of elements processed by the element model.

**4.4 Number of instructions vs. Number of Elements**

Getting the number of instructions executed during processing by the instruction model and comparing this result to the number of elements processed by the element model gives us a more accurate indication of the number of communication intervals and messages passed during execution.

Earlier in this dissertation, we said that if we have to improve the execution time and avoid less performance in parallel computing, we need ensure that the underlying parallel architecture produces the same results by reducing the number of instructions. Increasing the number of instructions means increased communication costs that have a critical impact on the overall performance of the architecture.

In order to obtain the number of instructions as well as the number of elements, a counter was introduced in both programs for this purpose. We looked at the most computational intensive part of the program and used a counter to keep track of every instruction executed as well as element processed. To understand this, we considered a simple formula that could also help compute the number of instructions using the assembly version of the code. The following formula which was used is:

$$S = in^3 + jn^2 + k + l$$

Where
$S$ represents the total number of instruction;
$i$ represents the number of instructions in the third loop of the program, this loop executes $n^3$ times;
$j$ represents the number of executed instructions in the first two loops, each loop executes $n$ times, hence $n^2$;
$n$ represents the size of the matrix (e.g. in a 100 x 100 matrix, $n$ is 100);
$k$ represents the number of instructions executed to generate the random number and all instructions executed at the beginning of the program and
$l$ represents the number of instructions executed after computation including all needed instructions to print the results as well as terminating the program.

As said earlier, we considered only the most intensive part of computation, which involves the loops. Although some instructions were left out for the purpose of making the analysis as simple as possible, the instruction model still produced a lot more instructions than AriDeM. Hence, the formula to count the number of instructions in the instruction model excluding $k$ and $l$ was refined to:

$$S = in^3 + jn^2$$

Where :
$S$ represents the total number of instruction;
$i$ represents the number of instructions in the third loop of the program, this loop executes $n^3$ times;
$j$ represents the number of executed instructions in the first two loops, each loop executes $n$ times, hence $n^2$;
$n$ represents the size of the matrix.

The approximate figures obtained for both models are indicated in table 5 below. The graph showing the progression in terms of these two factors: number of instructions versus number of elements for both models is given in figure 6.

Table 5
Von Neumann and AriDeM
(Number of instructions/elements per size of matrix)

|                    | 40 x 40  | 60 x 60  | 80 x 80   | 100 x100  |
|--------------------|----------|----------|-----------|-----------|
| Element Model      | 912000   | 3060000  | 7232000   | 14100000  |
| Instruction Model  | 2731200  | 9025200  | 21164800  | 41070000  |

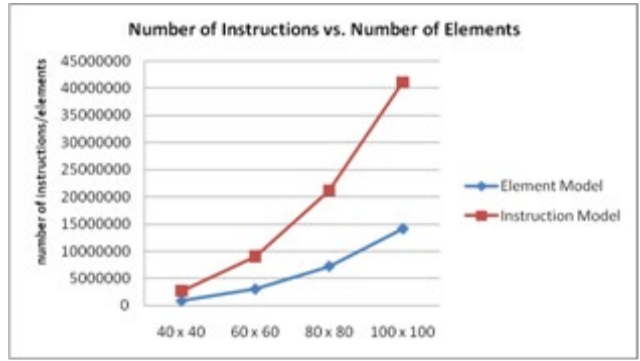

Figure 6: Progression of Number of instructions per size of matrix for AriDeM and von Neumann models

**Observation 5:**

Looking at the progression in figure 14, we notice some significant results for both models. The graph shows an indication of far better results obtained with AriDeM than von Neumann architecture. For the same matrix size, if computation takes place both models produce some kind of output. However, in order to do the same thing, AriDeM processes fewer elements compared to the number of instructions that are executed by the instruction model. As the graph indicates, the number of elements processed in AriDeM is almost three times less than the number of executed instructions with von Neumann.

Improving the processing time as well as the overall performance also heavily depends on the ability to reduce communication costs. The more instructions processors receive, the more messages need to be exchange between them and this leads to other performance costs such as overhead and result in delays. The empirical evidence has proved that AriDeM can perform the same task or computation as von Neumann but it can perform it while processing fewer elements.

When we have fewer instructions, we get fewer memory accesses. The number of instructions helps also determine the number of memory accesses. Although we did not specifically count the number of memory accesses during execution, the figures collected for the number of elements as well as the number of instructions give us an indication of the number of memory accesses.

## 5. SUMMARY AND CONCLUSION

In order to get these results needed to evaluate the instruction model, two separate programs were written to perform matrix multiplication. As a major justification for the model is to improve parallelism and the hardest case being tightly coupled, matrix multiplication was chosen to experiment with. The experiment required developing two programs: the matrix multiplication using the approach advocated in this paper and the simulation of the architecture writing in the address instruction based language C using MPI (Message Passing Interface). Both programs needed to be concurrent and be able to run on any number of processors. We chose to see the scalability of the models both with the number of processors as well as the size of matrices based on their execution time.

The experiment proved that a program like matrix multiplication could be written in the paradigm of the proposed model that can run on different configurations without any change and running it on 2, 4, 8 and 16 processors give the same results.

Despite the algorithms complexities, memory latency and other factors affecting the overall performance of the architectures, the results were enough to support the first findings and observations. The experiment gave an indication of how scalable these two models are. The instruction model could not show an effective synchronization of data processing as the results do not indicate scalability easily. Looking at the different graphs in this chapter, we can depict some discrepancies in terms of execution time and increasing size of matrix. We notice that as the size of matrix increases, the model does not strongly express that it is scalable. On the contrary, when the matrices 80 by 80 and 100 by 100 even 60 by 60 are run on 8 and 16 processors, we notice that the time actually goes up with 16 processors instead of decreasing. The same slightly happens for AriDeM but only with the 60 by 60 matrix. Otherwise, AriDeM distinguishes itself as being just a bit more scalable than von Neumann in this case.

In an attempt to get some comparison with a conventional architecture, we compared the number of instructions that need to be performed on the instruction architecture with the number of elements that need to be processed on this architecture. For the same matrix size, if computation takes place both models produce some kind of output. However, in order to do the same thing, AriDeM processes fewer elements compared to the number of instructions that are executed by the instruction model. As the graphs indicated, the number of elements processed in AriDeM is almost three times less than the number of executed instructions with von Neumann.The number of instructions generated shows that von Neumann needed almost three times the number of AriDeM elements to perform the same tasks.

As we have constantly emphasized throughout this dissertation, the experiment is a start to helping fine tune the model; identify weaknesses that need further investigation and potential strengths. While running the simulator on the conventional architecture, it gives us an indication of better performance with AriDeM. Whilst experimentation does show potential for the model, it is clear that this is just the start of the development of the model.

**BIBLIOGRAPHY OF AUTHORS (10 PT)**

| First author's Photo (3x4cm) | Patrick Mukala<br>He holds a master in software engineering from the Tshwane University of Technology, South Africa.<br>His research interests span from Software and Data Engineering to complex networks theory, information processing and Semantic Web. He is on the verge of completing a second master degree from the same institution. |
|---|---|
| | |